\documentclass[11pt,a4paper]{article}
\usepackage{cmap}
\usepackage[T1]{fontenc}
\usepackage[utf8]{inputenc}
\usepackage[margin=1in]{geometry}
\usepackage{amsmath,amssymb,bm}
\usepackage{graphicx}
\usepackage{booktabs,array}
\usepackage{cite}
\usepackage[labelfont=bf,font=small]{caption}
\usepackage{setspace}
\usepackage[colorlinks=true,linkcolor=blue,citecolor=blue,urlcolor=blue]{hyperref}
\graphicspath{{figs/}}
\newcommand{\dd}{\mathrm{d}}
\newcommand{\RD}{{\mathcal{R}_D}}
\newcommand{\ie}{\textit{i.e.}}

\title{\textbf{Cosmographic Analysis of Gravitational Baryogenesis in Running-$G$ Cosmology}}
\author{Nadeem Azhar$^{1}$\thanks{nadeemazharsaeed@gmail.com; nadeem.azhar@numl.edu.pk},
\  Abdul Jawad$^{2}$\thanks{jawadab181@yahoo.com;
abduljawad@ciitlahore.edu.pk}, \  Shamaila
Rani$^{2,~3}$\thanks{shamailatoor.math@yahoo.com;
drshamailarani@cuilahore.edu.pk}, \  Nurgissa
Myrzakulov$^{4}$\thanks{nmyrzakulov@gmail.com}, \\[4pt]
\  Sanjar Shaymatov$^{5,~6}$\thanks{sanjar@astrin.uz}\\[4pt]
{\small $^{1}$Department of Mathematics, National University of Modern Languages (NUML),}\\
{\small Lahore-Campus, Lahore-54000, Pakistan.}\\[2pt]
{\small $^{2}$Department of Mathematics, COMSATS Institute of Information Technology,}\\
{\small Lahore-54000, Pakistan.}\\[2pt]
{\small $^{3}$Research Center of Astrophysics and Cosmology, Khazar University,}\\
{\small Baku AZ1096, 41 Mehseti Street, Azerbaijan.}\\[2pt]
{\small $^{4}$L. N. Gumilyov Eurasian National University, Astana 010008, Kazakhstan.}\\[2pt]
{\small $^{5}$Institute of Fundamental and Applied Research,
National Research University TIIAME,}\\
{\small Kori Niyoziy 39, Tashkent 100000, Uzbekistan.}\\[2pt]
{\small$^{6}$University of Tashkent for Applied Sciences, Str.
Gavhar 1, Tashkent 100149, Uzbekistan.}}

\date{}
\begin{document}
\maketitle

\begin{abstract}
\noindent Gravitational baryogenesis produces a baryon asymmetry
through the coupling of the baryon current to the derivative of the
Ricci scalar, so its yield is set by the expansion history around
decoupling. We study this mechanism in cosmology with a running
Newton coupling of the asymptotic-safety type. Writing the Ricci
scalar and its derivative through the Hubble rate, the deceleration
parameter $q$ and the jerk $j$, we show that the yield is
proportional to the kernel $K=2+q-j$, which vanishes for radiation
in general relativity, and to a power of an enhancement factor
measuring how strongly the running coupling accelerates the
expansion at decoupling. We treat hybrid, intermediate, logamediate,
emergent, bouncing and little-rip scale factors and determine where
the observed asymmetry is reproduced for the $J^\mu\partial_\mu R$
and $J^\mu\partial_\mu R^2$ couplings; a table compares the six
cases.
\end{abstract}


\section{Introduction}
\label{sec:intro}

Everything we can observe on cosmological scales is made of matter.
Antimatter turns up only as a by-product of high-energy collisions,
and the net baryon excess is summarised by one small number, the
ratio of the baryon number density to the entropy density,
$\eta_B=n_B/s\simeq(8.70\pm0.04)\times10^{-11}$, which the cosmic
microwave background fixes with high precision \cite{Planck2018VI}.
An independent determination from the abundances of the light
elements formed during the first minutes gives a consistent value
\cite{Cyburt2016}. Why a Universe that is expected to have begun
with equal amounts of matter and antimatter ended with this
particular excess is the baryogenesis problem. Sakharov showed that
a dynamical generation requires three ingredients: violation of
baryon number, violation of $\mathcal C$ and $\mathcal{CP}$, and a
departure from thermal equilibrium \cite{Sakharov1967}. The Standard
Model contains all three in principle, but the $\mathcal{CP}$
violation of the quark sector is far too weak and the electroweak
transition is a smooth crossover, so a successful mechanism must go
beyond it.

The proposals that have been developed can be grouped by where the
three ingredients come from. In grand-unified scenarios, the
out-of-equilibrium decays of heavy gauge or Higgs bosons violate
baryon number directly \cite{Yoshimura1978,WeinbergPRL1979}. Since
anomalous electroweak processes (sphalerons) erase a baryon
asymmetry carried by left-handed fields at high temperature but
conserve the difference between baryon and lepton number
\cite{Kuzmin1985}, an asymmetry in lepton number produced earlier,
for instance in the decays of heavy right-handed neutrinos, can be
converted into a baryon asymmetry; this is leptogenesis
\cite{Fukugita1986}. Electroweak baryogenesis tries to use the
sphalerons together with a first-order phase transition, and
Affleck--Dine baryogenesis uses the coherent motion of a scalar
field that carries baryon number along flat directions of a
supersymmetric potential \cite{AffleckDine1985}. Detailed accounts
of these and related scenarios, with their strengths and
difficulties, are given in
Refs.~\cite{Riotto1999,Dine2003,Cline2006}. A conceptually different
route was identified by Cohen and Kaplan \cite{CohenKaplan1987}. If
a derivative coupling links the baryon current to a slowly varying
background field, particles and antiparticles acquire different
energies, exactly as if a chemical potential were present.
$\mathcal{CPT}$ is then broken dynamically by the background and the
third Sakharov condition is relaxed, since the asymmetry can be
generated while the plasma remains close to equilibrium. This
spontaneous mechanism needs a background that changes in time, and
the most economical candidate for it is the geometry of the
expanding Universe itself.

This is the idea behind gravitational baryogenesis, proposed by
Davoudiasl, et al.~\cite{Davoudiasl2004}. The interaction is
$M_\ast^{-2}\int\dd^4x\sqrt{-g}\,J^\mu\partial_\mu R$, where $J^\mu$
is the baryon current, $R$ the Ricci scalar and $M_\ast$ the cutoff
of the effective theory. In a spatially flat
Friedmann--Lema\^itre--Robertson--Walker (FLRW) Universe $\dot R$
acts as a chemical potential, and once the baryon-violating
processes fall out of equilibrium at the temperature $T_D$ the
frozen asymmetry is proportional to $\dot R(t_D)/(M_\ast^2T_D)$. In
general relativity, however, $R=(1-3w)\rho/M_{\rm P}^2$ for a fluid
with equation of state $w$, with $M_{\rm P}$ the reduced Planck
mass, so $R$ vanishes during radiation domination and the mechanism
is inactive there. The asymmetry is therefore a probe of departures
from the standard radiation-era expansion, and this has led to a
long list of studies in which the gravity sector is modified: $f(R)$
gravity \cite{Lambiase2006,Ramos2017}, Gauss--Bonnet gravity and its
braneworld form \cite{Odintsov2016,Bento2005}, $f(R,T)$ models
\cite{Baffou2019,Sahoo2020}, $f(T)$ and extended teleparallel
gravity \cite{Oikonomou2016,Bhattacharjee2020}, $f(R,L_m)$ gravity
\cite{Jaybhaye2023}, and models with quantum-fluctuation corrections
\cite{YangShi2024}. In every one of them the modification enters the
yield through the same channel, namely the expansion history near
$t_D$. This suggests treating the asymmetry as a cosmographic
observable, a function of the Hubble rate and of its first few
derivatives at decoupling, a point of view that we develop
quantitatively in this paper.

It is worth spelling out how the asymmetry arises in this class of
models, because the dependence on the background is what the rest of
the paper exploits. While the baryon-violating interactions are in
equilibrium, the effective chemical potential $\mu_B\propto\dot
R/M_\ast^2$ biases the equilibrium abundances of baryons and
antibaryons, so that $n_B\propto\mu_BT^2$. The asymmetry is erased
or preserved depending on whether these interactions are fast or
slow compared with the expansion, and the value that survives is the
one reached at the decoupling temperature $T_D$, when the
interaction rate drops below the Hubble rate. Dividing by the
entropy density, which scales as $g_\ast T^3$, removes the dilution
due to the subsequent expansion, and the ratio $n_B/s$ is conserved
thereafter if entropy is conserved. Three inputs therefore fix the
number: the cutoff $M_\ast$, the decoupling temperature $T_D$, and
the curvature derivative $\dot R$ at that moment. The first two are
conventionally taken near the scale of inflation and the reduced
Planck mass, and they cancel out of any comparison between different
expansion histories at fixed values. What remains is $\dot R(t_D)$
and the relation between $T_D$ and the expansion rate, which fixes
the epoch at which $\dot R$ is evaluated. Both are properties of the
background cosmology, and for this reason the same mechanism that in
general relativity fails during radiation domination can become
effective in any theory that changes the expansion law.

The gravity theories used in these studies have a common ancestry.
General relativity is not perturbatively renormalisable, as shown by
the one-loop calculation of 't~Hooft and Veltman
\cite{tHooftVeltman1974}. Adding quadratic curvature terms makes the
theory renormalisable at the price of ghosts \cite{Stelle1977},
whereas treating gravity as an effective field theory gives
controlled quantum corrections at low energy but no ultraviolet
completion \cite{Donoghue1994}. Higher-curvature terms were also
considered for cosmological reasons, the best-known case being the
$R^2$ model of Starobinsky \cite{Starobinsky1980}, and the general
family of $f(R)$ and related modified theories is reviewed in
Refs.~\cite{Sotiriou2010,Nojiri2011}. A different resolution was
proposed by Weinberg: gravity might be non-perturbatively
renormalisable if the renormalisation-group flow of its couplings
reaches a non-Gaussian ultraviolet fixed point, a scenario known as
asymptotic safety \cite{Weinberg1979}.

The functional renormalisation-group equation for the gravitational
effective average action \cite{Reuter1998} turned this proposal into
a calculable programme, and its status is reviewed in
Refs.~\cite{Niedermaier2006,Percacci2009,Litim2011,Eichhorn2019,Bonanno2020}.
The consequences for cosmology were soon explored. Bonanno and
Reuter built Planck-era cosmologies in which the running of $G$ and
of the cosmological term removes the initial singularity and leads
to a de Sitter phase \cite{BonannoReuter2002a,BonannoReuter2002b},
and Reuter and Saueressig constructed complete histories from the
big bang to asymptotic de Sitter space \cite{ReuterSaueressig2005}.
Later work addressed asymptotically safe inflation
\cite{Weinberg2010,Hindmarsh2011,BonannoPlatania2015}, the link
between the fixed point and dark energy \cite{Ahn2011}, and the
present status of the field \cite{BonannoSaueressig2017}. A
technical question in all of these is how to turn a scale-dependent
coupling into a cosmological equation of motion. The infrared scale
can be identified with a function of time, or promoted to a scalar
field so that the identification respects general covariance, and
the running couplings can be inserted at the level of the action or
of the field equations
\cite{Babic2005,ReuterWeyer2004,Domazet2012,Hindmarsh2012}. In
either case a time-dependent $G$ contributes a gradient term to the
effective stress--energy tensor and modifies the Friedmann
equations, which in turn changes the curvature that sources
gravitational baryogenesis.

Two features of the asymptotically safe description are important
for what follows. The first is the identification of the cutoff. In
the cosmological applications of Bonanno and Reuter the scale is
taken to be inversely proportional to the cosmic time, $k=\xi/t$
with $\xi$ a constant, so that the running coupling becomes an
explicit function of $t$: it grows as $t^2$ at early times, where
the fixed-point behaviour holds, and approaches the low-energy value
at late times. Different identifications, for instance with the
Hubble rate or with a curvature scale, lead to different histories,
which is one reason why the literature contains many cosmological
solutions of the same underlying flow. The second feature is the
logical status of the scale factor. In a complete treatment one
would solve the modified Friedmann equations together with the
running of $G$ and obtain $a(t)$; in practice the system is often
closed by assuming a power law for $a(t)$ and for $G(t)$ and
checking consistency with the conservation of the matter content.
The conclusions drawn from the second approach therefore depend on
the assumed functional form, and it is useful to know how much of
the baryogenesis result is a property of the form and how much is
generic.

Recently, Ref.~\cite{Yang2026} combined the two subjects and studied
gravitational baryogenesis in asymptotically safe gravity, using a
power-law scale factor $a\propto t^\alpha$ together with $G\propto
t^{\beta}$. It found that the observed asymmetry can be obtained
within narrow intervals of $\alpha$. That analysis raises questions
that the present paper addresses. First, the power law is an
assumption, and it is not clear which features of the result are due
to it and which are generic. Second, the yield depends on the
expansion history only through the curvature and its time derivative
at $t_D$, so it should be expressible in terms of the standard
cosmographic parameters
\cite{Visser2004,Sahni2003,Alam2003,Capozziello2019}, and doing so
would allow histories with very different global behaviour to be
compared on equal terms. Third, the literature on expansion laws
offers a variety of non-power-law scale factors, such as
intermediate \cite{Barrow2006} and logamediate
\cite{BarrowNunes2007} expansion, emergent universes
\cite{EllisMaartens2004}, bouncing cosmologies
\cite{Novello2008,Brandenberger2017} and little-rip scenarios
\cite{Frampton2011}, and it is natural to ask what each of them
implies for the asymmetry.

Our contributions are as follows. We summarise the running-$G$
cosmology and derive the linear equation that determines $G(t)$ for
a given scale factor and radiation-like matter
(Sec.~\ref{sec:cosmology}). We show that the Ricci scalar and its
derivative can be written as $R=6H^2(1-q)$ and $\dot
R=-6H^3(2+q-j)$, so that the baryon yield is governed by a single
kernel $K=2+q-j$ together with the expansion rate at decoupling,
which we characterise by one enhancement factor $\RD$
(Sec.~\ref{sec:GB}). We then work out hybrid, intermediate,
logamediate, emergent, bouncing and little-rip scale factors in
turn, giving for each the closed-form kinematics, the kernel and the
parameter region in which the observed asymmetry is reproduced for
two $\mathcal{CP}$-violating couplings (Sec.~\ref{sec:six}). The six
cases are compared in a table in Sec.~\ref{sec:compare}, and the
conclusions are given in Sec.~\ref{sec:conclusions}. We use units
$\hbar=c=k_B=1$, overdots for $\dd/\dd t$, primes for $\dd/\dd\ln
a$, and the metric signature $(-,+,+,+)$.

\section{Cosmology with a Running Newton Coupling}
\label{sec:cosmology}

In the asymptotic-safety scenario the gravitational couplings depend
on an energy scale $k$. Near the ultraviolet fixed point the
dimensionful Newton coupling behaves as $G_k\simeq g_\ast/k^2$, and
a commonly used interpolating form is $G_k=G_0/(1+\omega G_0k^2)$,
with $\omega$ a positive constant fixed by the flow
\cite{BonannoReuter2002a}. To use $G_k$ in a cosmological setting,
the scale $k$ is identified with a function of the cosmic time, or
more generally with a scalar field, so that the coupling becomes
$G(t)$ and $\Lambda(t)$ \cite{ReuterWeyer2004,Domazet2012}.
Implementing the scale dependence at the level of the action,
\begin{equation}
S[g]=\int\dd^4x\sqrt{-g}\left[\frac{R-2\Lambda_k}{16\pi G_k}+\mathcal L_m\right],
\label{eq:action}
\end{equation}
and varying with respect to the metric gives
\begin{equation}
G_{\mu\nu}=-\Lambda_kg_{\mu\nu}+8\pi G_kT_{\mu\nu}-\Delta t_{\mu\nu},\qquad
\Delta t_{\mu\nu}=G_k\left(g_{\mu\nu}\Box-\nabla_\mu\nabla_\nu\right)\frac{1}{G_k}.
\label{eq:EFE}
\end{equation}
The extra tensor $\Delta t_{\mu\nu}$ is the contribution of the
gradient of $G_k$ \cite{ReuterWeyer2004,Hindmarsh2012}, and it
vanishes when $G_k$ is constant.

For the spatially flat FLRW metric $\dd s^2=-\dd t^2+a^2(t)\,\dd\bm
x^2$ and a perfect fluid $T_{\mu\nu}=(\rho+p)u_\mu
u_\nu+pg_{\mu\nu}$ with $p=w\rho$, one has for a function $\phi(t)$
the identities $(g_{00}\Box-\nabla_0\nabla_0)\phi=3H\dot\phi$ and
$(\Box-\nabla^i\nabla_i)\phi=-\ddot\phi-2H\dot\phi$ (no sum over
$i$), with $H=\dot a/a$. The $00$ and $ii$ components of
Eq.~\eqref{eq:EFE} then give, writing $G\equiv G_k$ and
$\Lambda\equiv\Lambda_k$,
\begin{align}
H^2&=\frac{8\pi G}{3}\rho+\frac{\Lambda}{3}+H\frac{\dot G}{G},\label{eq:F1}\\
\frac{\ddot a}{a}&=-\frac{4\pi G}{3}(1+3w)\rho+\frac{\Lambda}{3}+\frac12\frac{\dot G}{G}H
+\frac12\frac{\ddot G}{G}-\left(\frac{\dot G}{G}\right)^2.\label{eq:F2}
\end{align}
Subtracting Eq.~\eqref{eq:F2} from Eq.~\eqref{eq:F1} removes the
cosmological term and, with $H^2-\ddot a/a=-\dot H$, leaves
\begin{equation}
-\dot H=4\pi(1+w)G\rho+\frac12\frac{\dot G}{G}H-\frac12\frac{\ddot G}{G}+\left(\frac{\dot G}{G}\right)^2 .
\label{eq:Hdot}
\end{equation}
The geometry can also be described by a total effective fluid,
defined with a reference coupling $G_0$ by $8\pi G_0\rho_{\rm
eff}\equiv3H^2$ and $8\pi G_0p_{\rm eff}\equiv-2\dot H-3H^2$. It
obeys $\dot\rho_{\rm eff}=-3H(\rho_{\rm eff}+p_{\rm eff})$
identically, and its equation of state is $w_{\rm eff}=-1-2\dot
H/(3H^2)$. The gradient terms of $G$ and the cosmological term are
part of this effective fluid, so $w_{\rm eff}$ generally differs
from the matter value $w=1/3$ of radiation. The difference between
the actual expansion rate and the one that radiation alone would
produce in general relativity is measured at decoupling by the
quantity $\RD$ introduced in Sec.~\ref{sec:GB}.

\subsection{Radiation era and the Equation for $G(t)$}

During the radiation era the matter obeys $w=1/3$ and the
conservation law gives $\rho=\rho_0a^{-4}$. Introducing
$u(t)\equiv1/G(t)$, so that $\dot G/G=-\dot u/u$ and $\ddot
G/G=-\ddot u/u+2(\dot u/u)^2$, Eq.~\eqref{eq:Hdot} multiplied by $u$
becomes a linear equation,
\begin{equation}
\ddot u-H\dot u+2\dot H\,u=-\frac{32\pi}{3}\,\rho_0\,a^{-4}.
\label{eq:uode}
\end{equation}
For any prescribed scale factor $a(t)$, Eq.~\eqref{eq:uode}
determines the running of $G$ up to the two integration constants of
the homogeneous equation, and $\Lambda(t)$ follows from
Eq.~\eqref{eq:F1}. Physical solutions require $u>0$.
Equation~\eqref{eq:uode} generalises the relation between the
exponents of the scale factor and of the coupling that is available
only for power laws. For $a=a_0(t/t_0)^\alpha$ and
$G=G_0(t/t_0)^\beta$ one finds
\begin{equation}
\rho=\frac{3}{32\pi}\,\frac{\mathcal N\,t_0^{\beta}}{G_0\,t^{\beta+2}},\qquad
\mathcal N\equiv2\alpha-\beta-\alpha\beta-\beta^2,\qquad \beta=4\alpha-2,
\label{eq:PLsol}
\end{equation}
where the last relation follows from $\rho\propto a^{-4}$.
Positivity of the density requires $\mathcal N>0$, which with
$\beta=4\alpha-2$ is
$(4-\sqrt6)/10<\alpha<(4+\sqrt6)/10\simeq0.645$. Because $\beta>0$
throughout the physically relevant range, $G$ grows with time, in
the direction expected from the antiscreening behaviour of the
fixed-point regime. In the other families treated below the function
$G(t)$ is not needed explicitly; as shown in Sec.~\ref{sec:GB} it
enters the baryon yield only through the single number $\RD$.

\subsection{Cosmographic Parameters}

Cosmological parameters are important because a scale factor is a
whole function, whereas observations constrain only its local
behaviour around a given epoch, and the coefficients of the Taylor
expansion of $a(t)$ are exactly the Hubble rate and the
dimensionless parameters built from its higher derivatives
\cite{Visser2004}. These numbers do not depend on a particular field
equation, so they give a common language for comparing general
relativity, modified gravity and running-coupling cosmology
\cite{Capozziello2019}, and they separate questions that are easily
confused: the Hubble rate sets the scale of the expansion, the
deceleration parameter says whether it is speeding up or slowing
down, and the jerk says how that tendency is itself changing. The
deceleration parameter indicates whether the expansion is slowing
down ($q>0$) or speeding up ($q<0$), with $q=0$ marking the switch
between the two phases. In a spatially flat universe governed by
general relativity it takes the value $1$ for radiation, $1/2$ for
pressureless matter and $-1$ for de Sitter expansion, while $q<-1$
means that the Hubble rate increases with time, $\dot H>0$.

The jerk parameter $j$ supplies information that $q$ cannot provide
on its own, since it determines how quickly $q$ changes. In a flat
universe filled with pressureless matter and a cosmological constant
($\Lambda$CDM) it equals unity at all times, so any deviation of $j$
from $1$ points to physics beyond that model
\cite{Sahni2003,Alam2003}. Moreover, $q = 1$ and $j = 3$ leads the
radiation dominant universe. It is convenient to describe the
expansion history by the dimensionless derivatives of the scale
factor. With $D_n\equiv a^{(n)}/a$, which obey $D_1=H$ and
$D_{n+1}=\dot D_n+HD_n$, the deceleration parameter and the jerk are
\cite{Visser2004}
\begin{equation}
q=-\frac{\ddot a\,a}{\dot a^2}=-1-\frac{\dot H}{H^2},\qquad
j=\frac{D_3}{H^3}=1+\frac{3\dot H}{H^2}+\frac{\ddot H}{H^3}=2q^2+q-q',
\label{eq:qj}
\end{equation}
where a prime denotes $\dd/\dd\ln a$. The jerk is the quantity
called $r$ in the statefinder pair \cite{Sahni2003,Alam2003}, and
the series of such parameters is the basis of cosmography in
modified gravity \cite{Capozziello2019}. Only $q$ and $j$ will be
needed below, because the baryon yield involves $R$ and $\dot R$;
higher parameters such as the snap would enter if the coupling
contained higher derivatives of $R$.

\section{Gravitational Baryogenesis}
\label{sec:GB}

The $\mathcal{CP}$-violating interaction is
\begin{equation}
\mathcal L_{\rm int}=\frac{1}{M_\ast^2}\,J^\mu\partial_\mu f(R),
\label{eq:Lint}
\end{equation}
with $f(R)=R$ in the original proposal and $f(R)=R^2/M_\ast^2$ as a
representative generalisation. In a homogeneous background only the
time component survives, $J^0\dot f/M_\ast^2$, which shifts the
energy of baryons and antibaryons in opposite directions. It acts
like a chemical potential $\mu_B=-\dot f(R)/M_\ast^2$ in the sign
convention of Ref.~\cite{Davoudiasl2004}. For relativistic fermions
with $\mu_B\ll T$ the net number density is $n_B=g_b\mu_BT^2/6$,
where $g_b$ counts the baryonic internal degrees of freedom, and the
entropy density is $s=2\pi^2g_\ast T^3/45$, with $g_\ast$ the number
of effectively massless species. Their ratio at the temperature
$T_D$ at which the baryon-violating interactions decouple is
\begin{equation}
\eta_B\simeq-\frac{15\,g_b}{4\pi^2g_\ast}\,\frac{\dot R\,f_R}{M_\ast^2\,T_D}
\bigg|_{t_D},\qquad f_R\equiv\frac{\dd f}{\dd R}.
\label{eq:etaB}
\end{equation}
For $f=R$ one has $f_R=1$, and for $f=R^2/M_\ast^2$ one has
$f_R=2R/M_\ast^2$, giving $\eta_B\simeq-30g_b\,R\dot R/(4\pi^2g_\ast
M_\ast^4T_D)$. We denote the two cases by the labels $(R)$ and
$(R^2)$.

For a flat FLRW metric $R=6(\dot H+2H^2)$. Using $\dot H=-(1+q)H^2$ one obtains
\begin{equation}
R=6H^2(1-q).
\label{eq:Rq}
\end{equation}
Differentiating $R$ and eliminating $\ddot H$ with the definition of
the jerk, $\ddot H=H^3(j-1)-3H\dot H$, gives
\begin{equation}
\dot R=6\big(\ddot H+4H\dot H\big)=6H^3\big[(j-1)-(1+q)\big]=-6H^3K,\qquad K\equiv2+q-j .
\label{eq:Rdot}
\end{equation}
We call $K$ the baryogenesis kernel. It vanishes for radiation in
general relativity ($q=1$, $j=3$), consistently with $R=0$ there,
and also for exact de Sitter expansion ($q=-1$, $j=1$), where $R$ is
constant. With Eqs.~\eqref{eq:etaB}, \eqref{eq:Rq} and
\eqref{eq:Rdot} the yield is
\begin{equation}
\eta_B^{(R)}=\frac{45\,g_b}{2\pi^2g_\ast}\,\frac{H_D^3\,K_D}{M_\ast^2T_D},\qquad
\eta_B^{(R^2)}=\frac{270\,g_b}{\pi^2g_\ast}\,\frac{H_D^5\,(1-q_D)\,K_D}{M_\ast^4T_D},
\label{eq:etaHK}
\end{equation}
where the subscript $D$ means evaluation at $t_D$. A positive $K_D$
gives a positive $\eta_B$, \ie\ an excess of baryons for the sign of
the coupling used here. The first expression shows that the
asymmetry depends on the expansion history only through $H_D$ and
the combination $2+q_D-j_D$; the line $j=2+q$ in the $(q,j)$ plane
is the locus of zero yield, and histories below it ($K>0$) produce
baryons while those above it ($K<0$) produce antibaryons.

We take, as in the earlier analysis \cite{Yang2026},
$T_D=M_I\simeq2\times10^{16}$~GeV, the scale of inflation allowed by
the tensor-mode bound \cite{Planck2018X}, $M_\ast=m_p/\sqrt{8\pi}$
with $m_p=1.22\times10^{19}$~GeV, $g_\ast=106$ and $g_b\simeq1$. The
radiation density at decoupling is $\rho_D=\pi^2g_\ast T_D^4/30$. If
the Universe were governed by general relativity with $G_0=1/m_p^2$,
the expansion rate would be $H_{\rm GR}^2=8\pi G_0\rho_D/3$, that is
$H_{\rm GR}=4.59\times10^{-5}\,m_p\simeq5.6\times10^{14}$~GeV. With
a running coupling the actual rate differs, and we parametrise the
difference by
\begin{equation}
\RD\equiv\frac{H_D^2}{H_{\rm GR}^2}=\frac{G_D}{G_0}\left[1+\frac{\Lambda_D+3H_D\dot G_D/G_D}{8\pi G_D\rho_D}\right],
\label{eq:RD}
\end{equation}
where the second equality follows from Eq.~\eqref{eq:F1}. The factor
$\RD$ collects everything that the running of $G$ does to the
expansion rate at decoupling: the change in the coupling itself, the
gradient term and the cosmological term. Inserting
$H_D=\sqrt{\RD}\,H_{\rm GR}$ in Eq.~\eqref{eq:etaHK} gives the
working formulas
\begin{equation}
\eta_B^{(R)}=3.20\times10^{-11}\,\RD^{3/2}\,K_D,\qquad
\eta_B^{(R^2)}=2.03\times10^{-17}\,\RD^{5/2}\,(1-q_D)\,K_D .
\label{eq:etaworking}
\end{equation}
Matching the observed value $8.70\times10^{-11}$ therefore requires
$\RD^{3/2}K_D=2.72$ for the $(R)$ coupling and
$\RD^{5/2}(1-q_D)K_D=4.3\times10^{6}$ for the $(R^2)$ coupling. The
second condition shows why the quadratic coupling needs a much
stronger enhancement: the extra factor $H_D^2/M_\ast^2$ is of order
$10^{-6}$ at $\RD=1$.

For a given scale-factor family the cosmographic state $(q_D,j_D)$
and hence $K_D$ is a function of the family's shape parameters and
of one ``phase'' variable that locates the decoupling time on the
expansion history, while $\RD$ is determined by the running of $G$.
The function $G(t)$ of a family follows from Eq.~\eqref{eq:uode}
only after the integration constants are specified, and it is not
unique. We therefore keep $\RD$ as a free parameter and report, for
every family, the value $\RD^{\rm req}$ that is needed to reproduce
$\eta_B^{\rm obs}$ together with the benchmark $\RD=4$, which is
close to the value $3.5$ that the closure of the power-law model
requires (see below).

\subsection{Reference Case: the Power-Law Solution}
\label{sec:PL} For $a\propto t^\alpha$ the parameters are constants,
$q=(1-\alpha)/\alpha$ and $j=(\alpha-1)(\alpha-2)/\alpha^2$, so that
\begin{equation}
K=\frac{2(2\alpha-1)}{\alpha^2},
\label{eq:Kpl}
\end{equation}
which vanishes for radiation ($\alpha=1/2$) and has the maximum
value $2$ at $\alpha=1$. In the power-law model the relations among
$\rho$, $G$ and $t$ close the system: the decoupling time follows
from $\rho(t_D)=\rho_D$ in Eq.~\eqref{eq:PLsol}, and
$\RD=3\alpha^2/(8\pi t_D^2\rho_D)$ in Planck units is then fixed.

\begin{figure}[ht]
\centering
\includegraphics[width=0.98\textwidth]{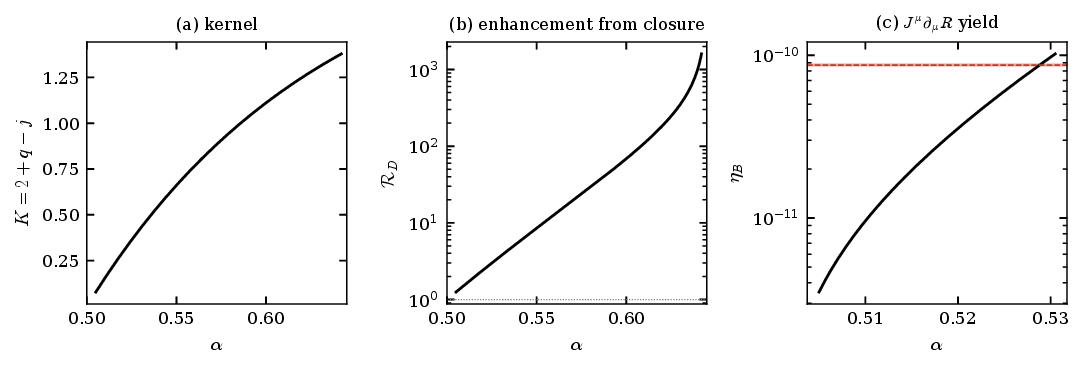}
\caption{Power-law reference model. (a) The kernel
$K=2(2\alpha-1)/\alpha^2$. (b) The enhancement $\RD$ obtained from
the closure of the model. (c) The yield $\eta_B^{(R)}$ of
Eq.~\eqref{eq:etaworking}; the dashed line and strip show the
observed value and its $1\sigma$ interval.} \label{fig:PL}
\end{figure}

Evaluating Eq.~\eqref{eq:etaworking} with this closure value is
algebraically equivalent to the yield formula of
Ref.~\cite{Yang2026}, and it agrees with the values tabulated there
to within 0.4\%, except for the row $\alpha=0.529$, where the
tabulated $8.63\times10^{-11}$ differs from our $8.86\times10^{-11}$
by 2.6\%; Table~\ref{tab:PL} gives the numbers, and
Fig.~\ref{fig:PL} the three ingredients. Matching $\eta_B^{\rm obs}$
fixes $\alpha=0.5288$ for the $(R)$ coupling and $\alpha=0.6355$ for
the $(R^2)$ coupling. The corresponding enhancement factors are
$\RD=3.5$ and $562$, and the coupling at decoupling is $G_D/G_0=2.7$
and $31.6$. These are the values with which the different scale
factors below will be compared. We note that the growth of $G$ by
one to two orders of magnitude between the Planck time and
decoupling, which the power-law ansatz implies, would have to stop
before the present epoch for the measured Newton constant to be
recovered; this is an inherent limitation of the ansatz.

\begin{table}[ht]
\centering \caption{Power-law reference model: kernel $K$,
decoupling time $t_D$ (units of $1/m_p$), enhancement $\RD$ from the
closure, coupling ratio $G_D/G_0$ and the yields for the two
couplings. Rows 3 and 6 are the observationally matched solutions.}
\label{tab:PL} \small
\begin{tabular}{@{}ccccccc@{}}
\toprule
$\alpha$ & $K$ & $t_D$ & $\RD$ & $G_D/G_0$ & $\eta_B^{(R)}$ & $\eta_B^{(R^2)}$\\
\midrule
0.5100 & 0.154 & 8883 & 1.56 & 1.44 & $9.60\times10^{-12}$ & $3.74\times10^{-19}$\\
0.5250 & 0.363 & 6606 & 2.99 & 2.41 & $6.01\times10^{-11}$ & $1.09\times10^{-17}$\\
0.5288 & 0.412 & 6137 & 3.52 & 2.73 & $8.69\times10^{-11}$ & $2.12\times10^{-17}$\\
0.5304 & 0.432 & 5951 & 3.77 & 2.88 & $1.01\times10^{-10}$ & $2.77\times10^{-17}$\\
0.6251 & 1.281 & 866 & 247 & 29.5 & $1.59\times10^{-7}$ & $1.00\times10^{-11}$\\
0.6355 & 1.342 & 584 & 562 & 31.6 & $5.71\times10^{-7}$ & $8.70\times10^{-11}$\\
0.6361 & 1.345 & 566 & 599 & 31.5 & $6.31\times10^{-7}$ & $1.03\times10^{-10}$\\
\bottomrule
\end{tabular}
\end{table}

\section{Baryogenesis for Different Scale Factors}
\label{sec:six}

For each family we give the scale factor, the Hubble rate and the
deceleration parameter, the cosmographic state at decoupling, the
kernel $K_D$, and the resulting yield. Throughout, $H_D$ denotes the
expansion rate at decoupling and $t_D$ the decoupling time; each
family has one dimensionless ``phase'' variable that fixes $t_D$ on
its expansion history for given $H_D$, and a few shape parameters.
Every closed expression for $q_D$, $j_D$ and $K_D$ was checked
against a direct symbolic differentiation of the scale factor. The
figures for each family have three panels at the same form: (a) the
kernel $K$, (b) the magnitude of the yield at the benchmark $\RD=4$
with the observed value shown as a dashed line and shaded strip, and
(c) the enhancement $\RD^{\rm req}$ needed to reproduce $\eta_B^{\rm
obs}$ with the $(R)$ coupling, drawn only where $K>0$. In (b), thick
lines are positive yields (baryon excess) and thin lines are the
magnitudes of negative ones.

\subsection{Hybrid Expansion}
\label{sec:hyb}
The hybrid law is a power law dressed with an
exponential factor, $a=a_0\,t^{\alpha}e^{\gamma t}$, with
$H=\alpha/t+\gamma$ and $\dot H=-\alpha/t^2$. For $\gamma t\ll1$ it
reduces to the reference model of Sec.~\ref{sec:PL}, and for $\gamma
t\gg1$ to de Sitter expansion. Define the phase variable
$\epsilon\equiv\gamma/H_D\in[0,1)$ and $X_D\equiv\alpha+\gamma
t_D=\alpha/(1-\epsilon)$. Then
\begin{equation}
q_D=-1+\frac{(1-\epsilon)^2}{\alpha},\qquad
j_D=1-\frac{3(1-\epsilon)^2}{\alpha}+\frac{2(1-\epsilon)^3}{\alpha^2},\qquad
K_D=\frac{2(1-\epsilon)^2\,(2\alpha-1+\epsilon)}{\alpha^2}.
\label{eq:Khyb}
\end{equation}

At $\epsilon=0$ the kernel is that of Eq.~\eqref{eq:Kpl}. Increasing
$\epsilon$ first raises $K_D$, since the exponential factor adds to
$\dot R$, up to a maximum $K_{\max}=64\alpha/27$ at
$\epsilon=1-4\alpha/3$ (valid for $\alpha<3/4$), and then lowers it
to zero as the expansion becomes de Sitter-like at $\epsilon\to1$.
For $\alpha>1/2$ the kernel is positive for all $\epsilon<1$, so the
hybrid law always produces baryons. Since $K_{\max}$ lies between
1.21 and 1.52 for $\alpha=0.51$--$0.64$, the $(R)$ coupling cannot
reproduce $\eta_B^{\rm obs}$ at $\RD=1$, which needs $K_D\geq2.72$.
At the benchmark $\RD=4$ the required kernel is $K_D=0.34$, which is
reached for $\alpha=0.55$ at $\epsilon=0.755$ (where $q_D=-0.89$ and
$j_D=0.77$) and for $\alpha=0.51$ at both $\epsilon=0.027$ and
$\epsilon=0.762$. These solutions show two distinct regimes: for
$\alpha$ very close to $1/2$ a small exponential rate suffices
because $2\alpha-1$ is small and $\epsilon$ dominates the bracket in
Eq.~\eqref{eq:Khyb}; otherwise the matching state lies close to de
Sitter expansion, with the exponential factor providing three
quarters of the Hubble rate. The smallest enhancement that works is
$\RD^{\rm req}=1.48$--$1.72$ for the $(R)$ coupling and about $400$
for the $(R^2)$ coupling, which is close to the values $3.5$ and
$562$ of the reference model.

\begin{figure}[ht]
\centering
\includegraphics[width=0.98\textwidth]{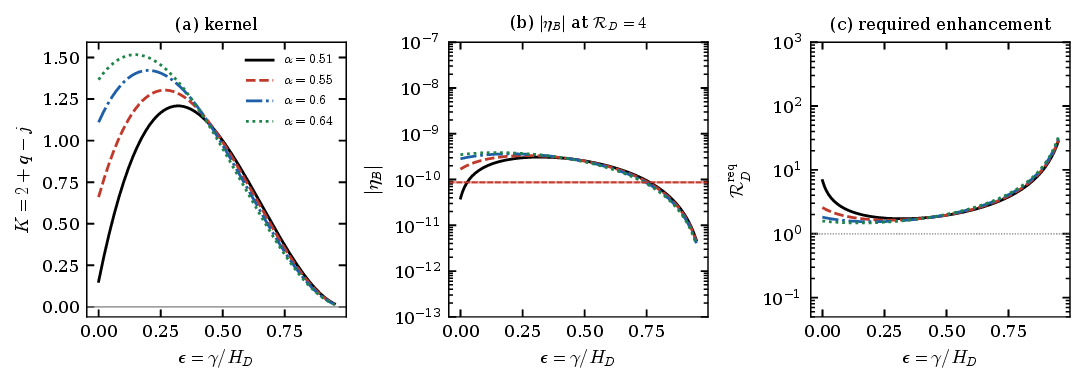}
\caption{Hybrid expansion. (a) Kernel $K$ as a function of
$\epsilon=\gamma/H_D$ for $\alpha=0.51,0.55,0.60,0.64$. (b) Yield
$|\eta_B^{(R)}|$ at $\RD=4$. (c) Enhancement $\RD^{\rm req}$ needed
to match $\eta_B^{\rm obs}$.} \label{fig:hyb}
\end{figure}

\subsection{Intermediate Expansion}
\label{sec:int}
The intermediate law is $a=a_0\exp(At^f)$ with
$0<f<1$, so that $H=Aft^{f-1}$ and $\dot H=-(1-f)H/t$. It expands
faster than any power law but slower than exponentially. The phase
variable is $x\equiv H_Dt_D=Aft_D^f$, and
\begin{equation}
q_D=-1+\frac{1-f}{x},\qquad
q'_D=-\frac{f(1-f)}{x^2},\qquad
K_D=\frac{1-f}{x^2}\,\big(4x-2+f\big).
\label{eq:Kint}
\end{equation}

\begin{figure}[ht]
\centering
\includegraphics[width=0.98\textwidth]{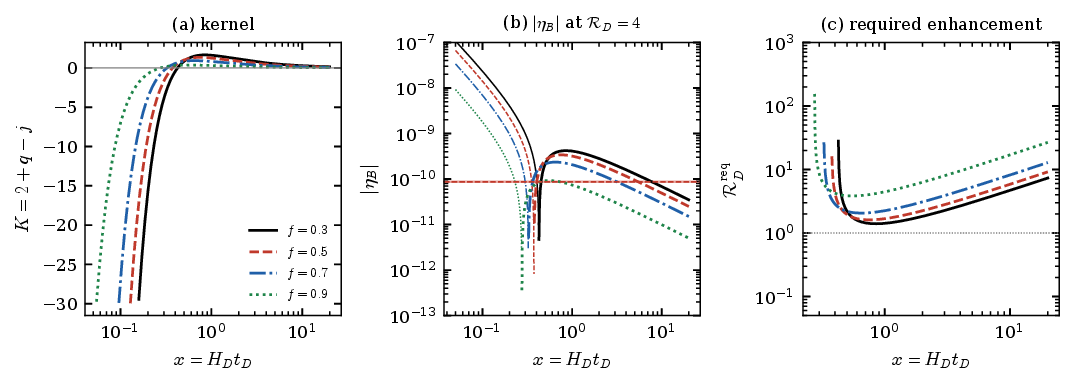}
\caption{Intermediate expansion for $f=0.3,0.5,0.7,0.9$ as a
function of $x=H_Dt_D$. Panels as in Fig.~\ref{fig:hyb}.}
\label{fig:int}
\end{figure}

At small $x$ the expansion is strongly decelerated, $q_D\gg1$, and
the kernel is large and negative (an antibaryon excess); $K_D$
changes sign at $x=(2-f)/4$, where $q_D=(2-3f)/(2-f)$, and reaches
the maximum $K_{\max}=4(1-f)/(2-f)$ at $x=(2-f)/2$, after which it
decays as $4(1-f)/x$ towards de Sitter. The maximum falls from 1.65
at $f=0.3$ to 0.36 at $f=0.9$, as expected since $f\to1$ is exact de
Sitter expansion with $K=0$. Consequently the required enhancement
increases with $f$, from $\RD^{\rm req}=1.40$ at $f=0.3$ to $3.83$
at $f=0.9$, and the $(R^2)$ coupling needs between 345 and 530. For
$f=0.5$ the observed asymmetry at $\RD=4$ is reproduced at $x=0.40$
(a still decelerated state with $q_D=0.24$ and $j_D=1.90$) and at
$x=5.5$ (near de Sitter, $q_D=-0.91$, $j_D=0.75$); the two roots are
the two sides of the maximum of $K_D$ in Fig.~\ref{fig:int}(a).

\subsection{Logamediate Expansion}
\label{sec:log}

The logamediate law is $a=a_0\exp\!\big[A\ln^\lambda(t/t_s)\big]$
with $\lambda>1$ and a scale $t_s$. With $L\equiv\ln(t/t_s)$ one
finds $H=A\lambda L^{\lambda-1}/t$ and $\dot
H=-H\,[1-(\lambda-1)/L]/t$. We use $x\equiv H_Dt_D=A\lambda
L_D^{\lambda-1}$ as the phase variable, and keep $L_D=4$ fixed as a
shape parameter so that $\lambda$ is the only remaining one. Writing
$c\equiv1-(\lambda-1)/L_D$,
\begin{equation}
q_D=-1+\frac{c}{x},\qquad
q_D'=\frac{(\lambda-1)}{L_Dx^2}\left(\frac1{L_D}-c\right),\qquad
K_D=\frac{4c}{x}-\frac{2c^2-(\lambda-1)(1/L_D-c)/L_D}{x^2}.
\label{eq:Klog}
\end{equation}
For $\lambda=1$ one has $c=1$ and $x=A$, so the law is the power law
$a\propto t^A$, and $K_D=2(2x-1)/x^2$ reproduces Eq.~\eqref{eq:Kpl}.

The behaviour of $K_D$ resembles that of the intermediate law:
negative for small $x$, a maximum, and a decay to zero as $x$ grows.
The maxima for $\lambda=1.5,2,3,4$ are $1.90$, $1.80$, $1.60$ and
$2.00$, and the corresponding minimum enhancements are $\RD^{\rm
req}=1.27$, $1.32$, $1.43$ and $1.23$ for the $(R)$ coupling, with
$340$--$346$ for the $(R^2)$ coupling. These are the smallest values
among the families that stay in the neighbourhood of power-law-like
expansion. For $\lambda=2$ the benchmark $\RD=4$ is matched at
$x=0.44$ ($q_D=0.71$, $j_D=2.37$) and at $x=8.4$ ($q_D=-0.91$,
$j_D=0.75$). As for the intermediate law, the maximum kernel is
below $2.72$, so $\RD=1$ cannot reproduce the observed asymmetry.

\begin{figure}[ht]
\centering
\includegraphics[width=0.98\textwidth]{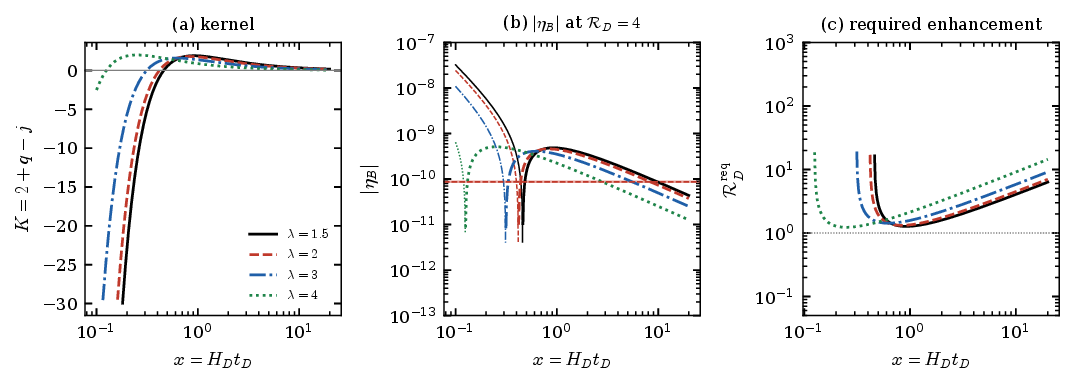}
\caption{Logamediate expansion with $L_D=4$ and $\lambda=1.5,2,3,4$,
as a function of $x=H_Dt_D$. Panels as in Fig.~\ref{fig:hyb}.}
\label{fig:log}
\end{figure}

\subsection{Emergent Expansion}
\label{sec:eme}

The emergent law is $a=a_i\,(1+e^{\zeta t})^n$
\cite{EllisMaartens2004}: the Universe approaches a static state in
the past and expands as $e^{n\zeta t}$ later. It has $H=n\zeta
s/(1+s)$ with $s\equiv e^{\zeta t}$, and $\dot H>0$, so it is
super-accelerated with $q<-1$ at all times. With $s_D=e^{\zeta t_D}$
as the phase variable and $u\equiv1/(ns_D)$,
\begin{equation}
q_D=-1-u,\qquad
j_D=1+3u+u^2-\frac{u}{n},\qquad
K_D=u\left(\frac1n-4-u\right).
\label{eq:Keme}
\end{equation}

The kernel is positive only if $u<1/n-4$, which requires $n<1/4$ and
$s_D>1/(1-4n)$; for $n\geq1/4$ it is negative for all $s_D$ and the
yield is a net antibaryon excess. For $n<1/4$ the maximum is
$K_{\max}=(1/n-4)^2/4$ at $s_D=2/(1-4n)$. This is large for small
$n$: $K_{\max}=9$ for $n=0.1$ and $0.25$ for $n=0.2$. For $n=0.1$
the maximum exceeds $2.72$, so the $(R)$ coupling can reproduce the
observed asymmetry already at $\RD<1$; the smallest enhancement is
$\RD^{\rm req}=0.45$, and $98$ for the $(R^2)$ coupling. The matched
state at $\RD=4$ is $s_D=1.68$ with $q_D=-6.9$ and $j_D=-5.3$, \ie\
far from radiation-like expansion. For $n=0.2$ the window is narrow
($\RD^{\rm req}=4.9$), and for $n=0.5$ and $1$ no matching with a
positive yield exists.

\begin{figure}[ht]
\centering
\includegraphics[width=0.98\textwidth]{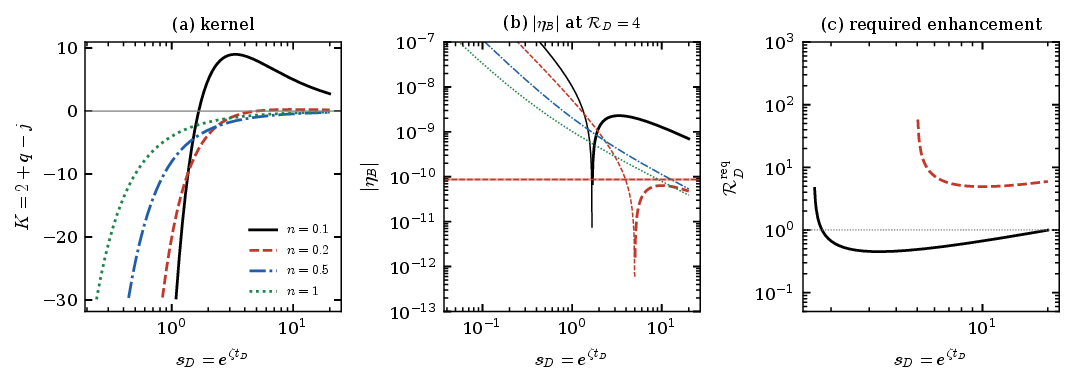}
\caption{Emergent expansion for $n=0.1,0.2,0.5,1$ as a function of
$s_D=e^{\zeta t_D}$. Panels as in Fig.~\ref{fig:hyb}; for $n\geq0.5$
the kernel is negative throughout.} \label{fig:eme}
\end{figure}

\subsection{Bouncing Expansion}
\label{sec:bou}

The bouncing law is $a=a_0(1+bt^2)^n$, with $H=2nbt/(1+bt^2)$, a
bounce at $t=0$ and the power-law behaviour $a\propto t^{2n}$ at
late times. We use $z_D\equiv bt_D^2$ on the expanding branch $t>0$
as the phase variable. Then $H_Dt_D=2nz_D/(1+z_D)$ and
\begin{equation}
q_D=-1+\frac{1}{2n}\left(1-\frac{1}{z_D}\right),\qquad
K_D=\frac{2(z_D-1)}{n\,z_D}+\frac{3-z_D}{2n^2z_D}.
\label{eq:Kbou}
\end{equation}

Near the bounce ($z_D\to0$) the kernel behaves as
$K_D\simeq(3-4n)/(2n^2z_D)$: it diverges to $+\infty$ for $n<3/4$
and to $-\infty$ for $n>3/4$, because $H\to0$ and $q\to-\infty$
there. At late times $K_D\to(4n-1)/(2n^2)$, which is the power-law
value of Eq.~\eqref{eq:Kpl} with $\alpha=2n$; in particular $n=1/4$
is asymptotically radiation, with $K\to0$ from above. A positive and
arbitrarily large yield is therefore obtainable just after the
bounce for $n<3/4$, which is why the enhancement required is the
smallest of all the families: $\RD^{\rm req}=0.105$, $0.372$ and
$1.33$ for $n=0.25$, $0.5$ and $0.75$, evaluated at $z_D=0.2$ (the
edge of the plotted range, so these are upper bounds on the
minimum), and $1.50$ for $n=1$, where the maximum $K_D$ occurs at
late times. The corresponding $(R^2)$ values are $31$, $81$, $195$
and $325$. For $n=0.25$ the benchmark is matched at $z_D=47$
($q_D=0.96$, $j_D=2.6$), a state close to radiation expansion
approached from above, and for $n=1$ at $z_D=0.43$. We note that
decoupling at $z_D\lesssim1$ means that the baryon-violating
processes freeze out within a Hubble time of the bounce, which is a
strong requirement on the thermal history of the bouncing model.

\begin{figure}[ht]
\centering
\includegraphics[width=0.98\textwidth]{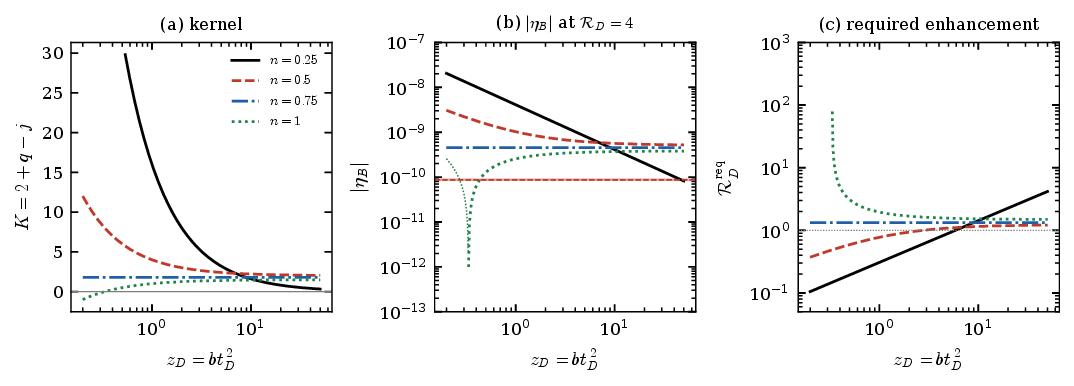}
\caption{Bouncing expansion for $n=0.25,0.5,0.75,1$ as a function of
$z_D=bt_D^2$. Panels as in Fig.~\ref{fig:hyb}.} \label{fig:bou}
\end{figure}

\subsection{Little-rip Expansion}
\label{sec:lrip}

In the little-rip scenario \cite{Frampton2011} the Hubble rate grows
exponentially, $H=H_0e^{\lambda t}$, so that
$a=a_0\exp\!\big[(H_0/\lambda)(e^{\lambda t}-1)\big]$ and $\dot
H=\lambda H>0$. Define $\epsilon\equiv\lambda/H_D$. Then
\begin{equation}
q_D=-1-\epsilon,\qquad j_D=1+3\epsilon+\epsilon^2,\qquad K_D=-4\epsilon-\epsilon^2.
\label{eq:Klrip}
\end{equation}
The kernel is negative for every $\epsilon>0$, so with the sign
convention of Eq.~\eqref{eq:Lint} the little-rip expansion produces
an antibaryon excess, and likewise for the $(R^2)$ coupling since
$1-q_D=2+\epsilon>0$. The yield magnitude can still match the
observed value: $|\eta_B^{(R)}|=\eta_B^{\rm obs}$ at
$\epsilon=0.593$, $0.083$ and $0.011$ for $\RD=1$, $4$ and $16$, so
a small rate $\lambda\ll H_D$ is enough. Because the sign is fixed
by the geometry ($\dot H>0$), the little-rip family cannot account
for a matter excess unless the sign of the coupling in
Eq.~\eqref{eq:Lint} is reversed, which would at the same time
reverse the sign of the yield in all the other families of this
paper.

\begin{figure}[ht]
\centering
\includegraphics[width=0.98\textwidth]{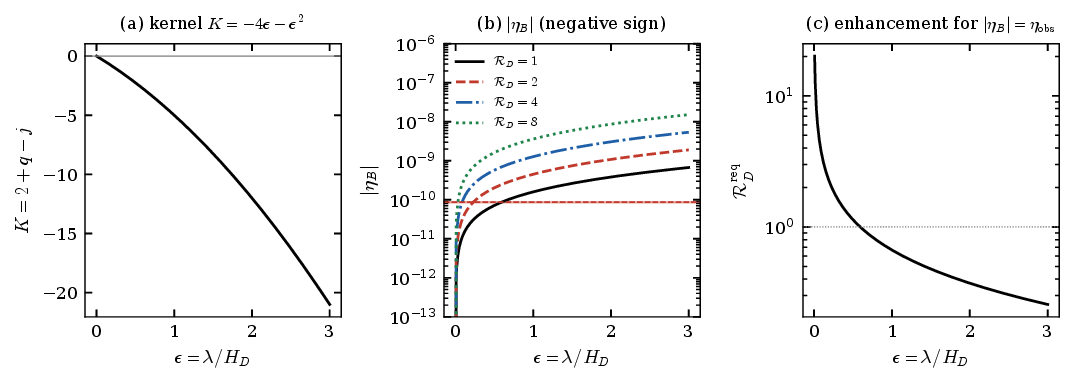}
\caption{Little-rip expansion as a function of
$\epsilon=\lambda/H_D$. (a) Kernel $K=-4\epsilon-\epsilon^2$. (b)
Magnitude of the (negative) yield for $\RD=1,2,4,8$. (c) Enhancement
for which the magnitude equals $\eta_B^{\rm obs}$.} \label{fig:lrip}
\end{figure}

\section{Comparison of the Scale Factors}
\label{sec:compare}

\begin{figure}[ht]
\centering
\includegraphics[width=0.98\textwidth]{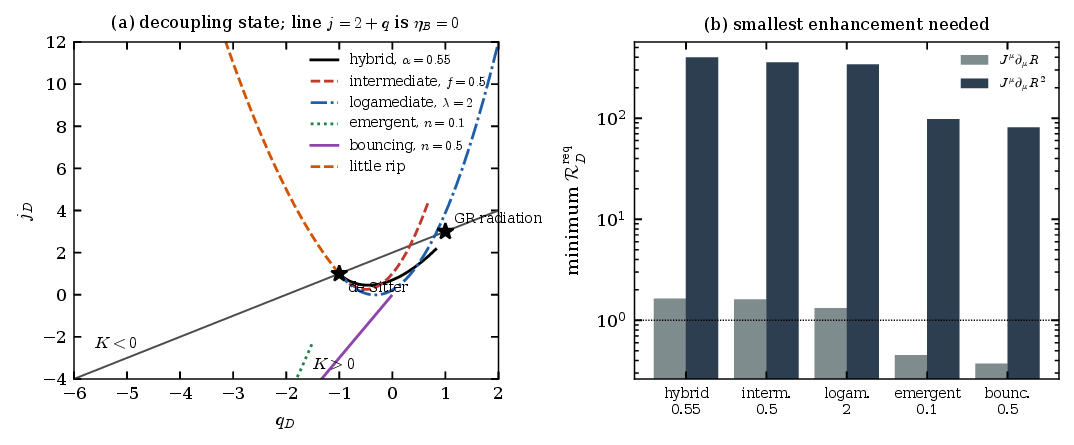}
\caption{(a) Decoupling states $(q_D,j_D)$ traced by one member of
each family as its phase variable is varied; the straight line
$j=2+q$ separates baryon production ($K>0$, below the line) from
antibaryon production ($K<0$). The stars mark radiation in general
relativity and de Sitter space, both on the line. (b) Smallest
enhancement $\RD^{\rm req}$ for which the observed asymmetry is
reproduced, for the two couplings; values below $1$ mean a weaker
expansion rate than in general relativity is required.}
\label{fig:comp}
\end{figure}

Figure~\ref{fig:comp} shows the decoupling states $(q_D,j_D)$ of
representative members of the six families against the zero-yield
line $j=2+q$, together with the smallest enhancement each family
needs. Table~\ref{tab:comp} collects the results. The kernel is a
function of the family and of the decoupling phase, so the
comparison is made at the point of largest kernel, where the
enhancement needed is smallest, and at the benchmark $\RD=4$.

\begin{table}[ht]
\centering \caption{Comparison of the scale factors. $K_{\max}$ is
the maximum kernel with respect to the phase variable (for the
bouncing law it diverges at the bounce and the value quoted is at
$z_D=0.2$ for $n\le0.75$). $\RD^{\rm req}$ is the enhancement needed
for $\eta_B^{\rm obs}$ with the $(R)$ coupling at $K_{\max}$, and
$\RD^{\rm req}_2$ the same for the $(R^2)$ coupling. The last column
states whether $\RD=1$ is sufficient for the $(R)$ coupling, which
needs $K_D\geq2.72$. The reference power-law values are those of
Table~\ref{tab:PL}.} \label{tab:comp} \footnotesize
\renewcommand{\arraystretch}{1.3}
\setlength{\tabcolsep}{3pt}
\newcolumntype{L}[1]{>{\raggedright\arraybackslash}p{#1}}
\begin{tabular}{@{}L{2.0cm}L{2.5cm}L{3.0cm}L{1.6cm}L{1.9cm}L{1.7cm}L{1.2cm}@{}}
\toprule
Scale factor & Kernel $K_D$ & $K_{\max}$ (parameters) & Sign of $\eta_B$ & $\RD^{\rm req}$ ($R$) & $\RD^{\rm req}_2$ ($R^2$) & $\RD=1$ enough\\
\midrule
Power law (ref.) & $\dfrac{2(2\alpha-1)}{\alpha^2}$ & $<1.35$ for $\alpha<0.645$ & $+$ & $3.5$ (closure) & $562$ (closure) & no\\
Hybrid $t^\alpha e^{\gamma t}$ & Eq.~\eqref{eq:Khyb} & $64\alpha/27$ (1.21--1.52) & $+$ & 1.48--1.72 & $\approx400$ & no\\
Intermediate $e^{At^f}$ & Eq.~\eqref{eq:Kint} & $\dfrac{4(1-f)}{2-f}$ (1.65--0.36) & $-$ for $x<\tfrac{2-f}{4}$, else $+$ & 1.40--3.83 & 345--530 & no\\
Logamediate $e^{A\ln^\lambda t}$ & Eq.~\eqref{eq:Klog} & 1.60--2.00 ($\lambda=1.5$--4, $L_D=4$) & $-$ small $x$, else $+$ & 1.23--1.43 & 340--346 & no\\
Emergent $(1+e^{\zeta t})^n$ & $u(1/n-4-u)$ & $(1/n-4)^2/4$ for $n<\tfrac14$ (9 at $n=0.1$) & $+$ only for $n<\tfrac14$ & 0.45 ($n=0.1$); 4.9 ($n=0.2$) & 98; 542 & yes for $n=0.1$\\
Bouncing $(1+bt^2)^n$ & Eq.~\eqref{eq:Kbou} & 80, 12, 1.78 ($n=0.25,0.5,0.75$); 1.49 ($n=1$) & $+$ for $n<\tfrac34$ near bounce & 0.105--1.50 & 31--325 & yes for $n\le0.5$\\
Little rip $e^{(H_0/\lambda)(e^{\lambda t}-1)}$ & $-4\epsilon-\epsilon^2$ & none ($K<0$) & $-$ & $|\eta|$ matched at $\epsilon=0.59$ ($\RD=1$) & --- & sign wrong\\
\bottomrule
\end{tabular}
\end{table}

The table shows the following pattern. Families that remain close to
a power-law-like decelerated or slowly accelerating expansion
(hybrid, intermediate, logamediate) have kernels bounded by about 2
and need enhancement factors of order 1.2--4 for the $(R)$ coupling
and a few hundred for the $(R^2)$ coupling, close to the values of
the reference model. The families with strong departures from the
standard expansion, namely the emergent law with $n<1/4$ and the
bouncing law near the bounce, have unbounded or large kernels and
can reproduce the asymmetry with $\RD\lesssim1$. The little-rip law,
with $\dot H>0$ at all times, has a kernel of fixed negative sign.

\section{Conclusions}
\label{sec:conclusions}

We have studied gravitational baryogenesis in cosmology with a
running Newton coupling of the asymptotic-safety type, with the aim
of understanding how the baryon asymmetry depends on the expansion
history, and we have worked out six scale factors in detail. The
main results are as follows.

The first result is structural. After summarising the field
equations of running-$G$ cosmology and the linear equation,
Eq.~\eqref{eq:uode}, that determines $G(t)$ for a prescribed scale
factor in the radiation era, we showed that the Ricci scalar and its
derivative can be written exactly as $R=6H^2(1-q)$ and $\dot
R=-6H^3(2+q-j)$. The baryon yield is therefore a cosmographic
observable: for the $J^\mu\partial_\mu R$ coupling it is
proportional to $H_D^3K_D$ with the kernel $K=2+q-j$, and for the
quadratic coupling to $H_D^5(1-q_D)K_D$. The kernel vanishes for
radiation in general relativity and for de Sitter expansion, and its
sign decides whether baryons or antibaryons are produced. All the
effects of the running of $G$ on the expansion rate at decoupling
are summarised in one number $\RD=H_D^2/H_{\rm GR}^2$, and the
yields scale as $\RD^{3/2}$ and $\RD^{5/2}$. With this formulation,
we recovered the published power-law results to within 0.4\% (one
tabulated row differs by 2.6\%), and we found that the matched
power-law solutions correspond to $\RD=3.5$ for the standard
coupling and $\RD=562$ for the quadratic one.

The second result concerns the families whose kinematics stays close
to a power law. For the hybrid law the kernel has the closed form
$K=2(1-\epsilon)^2(2\alpha-1+\epsilon)/\alpha^2$ with
$\epsilon=\gamma/H_D$, a maximum $64\alpha/27$, and positive sign
for all $\alpha>1/2$. For the intermediate and logamediate laws the
kernel is negative in the strongly decelerated regime, positive near
a maximum of $4(1-f)/(2-f)$ in the intermediate case, and decays to
zero as the expansion approaches de Sitter. All three have maximum
kernels below about 2 and need $\RD$ between 1.2 and 4 with the
standard coupling, and about 340 to 530 with the quadratic coupling,
which is the same range as for the reference model. With any of them
the observed asymmetry is reproduced in two kinds of state: a
decelerated one, or a nearly de Sitter one with a jerk somewhat
below unity.

The third result concerns the families with strong departures from
standard expansion. The emergent law, which is super-accelerated at
all times, gives a positive yield only for $n<1/4$, with
$K_{\max}=(1/n-4)^2/4$, and for $n=0.1$ the observed asymmetry is
obtained with $\RD<1$. The bouncing law has a kernel that diverges
just after the bounce for $n<3/4$ and tends to the power-law value
at late times, so for $n\le0.5$ the asymmetry can be obtained with a
weaker expansion rate than in general relativity, at the cost of
requiring decoupling within a Hubble time of the bounce. The
little-rip law has the kernel $-4\epsilon-\epsilon^2$ and produces
antibaryons for the sign of the coupling used here; since $\dot H>0$
fixes the sign, it cannot supply a matter excess in this scenario,
whatever the parameters.

The fourth result is the comparison in Table~\ref{tab:comp} and
Fig.~\ref{fig:comp}. The yield depends on the family chiefly through
the kernel and on the running of $G$ through $\RD$, and a family is
viable if its kernel reaches the required value
$K=2.72\,\RD^{-3/2}$. With the quadratic coupling the required
enhancement is of order $10^2$ to $10^3$ for all the families that
stay near a power law, and about 30 to 100 for the bouncing
($n\le0.5$) and emergent ($n=0.1$) cases. This means that the
quadratic coupling relies even more strongly than the standard one
on a large growth of $G$ before decoupling.

Several limitations should be kept in mind. First, we did not
construct $G(t)$ for the non-power-law families, since the
radiation-era equation has a two-parameter family of solutions for
each scale factor; $\RD$ was therefore treated as a free parameter
and the results should be read as requirements on it, not as
predictions. Obtaining $\RD$ from a definite renormalisation-group
trajectory is the natural next step, and it would convert the
required values listed here into tests of specific fixed-point
scenarios. Second, in the power-law reference model the matching
values of $\RD$ and $G_D/G_0$ are far from unity, in particular for
the quadratic coupling ($G_D/G_0\simeq32$), and the growth of $G$
would have to stop before the present epoch; scenarios in which $G$
saturates after the radiation era are more realistic and would need
to be examined. Third, we used the standard decoupling temperature
$T_D=2\times10^{16}$~GeV and cutoff $M_\ast=m_p/\sqrt{8\pi}$ for all
families so as to isolate the effect of the scale factor; other
values shift the required enhancement, since the yield scales as
$H_D^3/(M_\ast^2T_D)$. Finally, the asymmetry was obtained by
choosing parameters, so a match with the observed value is a
statement about the parameter space of each model and not an
independent confirmation of the mechanism. Extensions of this work
include a perturbation analysis of the effective fluid, other
$\mathcal{CP}$-violating couplings that involve $\ddot R$ and
therefore the snap parameter, and the reconstruction of $G(t)$ for
the families treated here.

\end{document}